\documentclass[12pt]{article}
\usepackage{graphicx} 
\usepackage{upquote} 
\usepackage{authblk}

\title{A curated UK rain radar data set for training and benchmarking nowcasting models}
\author[1]{Viv Atureta}
\author[1]{Rifki Priansyah Jasin}
\author[1]{Stefan Siegert\footnote{Corresponding author: S. Siegert, Department of Mathematics and Statistics, University of Exeter, United Kingdom, Email: \texttt{s.siegert@exeter.ac.uk}}}
\affil[1]{University of Exeter, Exeter, United Kingdom}
\date{\today}

\usepackage{xcolor}

\usepackage{enumitem}
\setlist{nosep}

\usepackage{booktabs}

\usepackage[T1]{fontenc}
\usepackage{hyperref}

\usepackage[most]{tcolorbox}
\tcbset{colback=gray!10, colframe=gray!60, boxrule=0.5pt, arc=2pt}

\usepackage{float}
\usepackage{placeins}

\begin{document}

\maketitle

\begin{tcolorbox}
\begin{center}
{\bf The data and code described in this paper can be downloaded at\\
\url{https://dx.doi.org/10.5281/zenodo.17713617}}
\end{center}
\end{tcolorbox}

\section*{Abstract}

This paper documents a data set of UK rain radar image sequences for use in statistical modeling and machine learning methods for nowcasting.
The main dataset contains 1,000 randomly sampled sequences of length 20 steps (15-minute increments) of 2D radar intensity fields of dimension 40x40 (at 5km spatial resolution). 
Spatially stratified sampling ensures spatial homogeneity despite removal of clear-sky cases by threshold-based truncation.
For each radar sequence, additional atmospheric and geographic features are made available, including date, location, mean elevation, mean wind direction and speed and prevailing storm type.
New R functions to extract data from the binary "Nimrod" radar data format are provided.
A case study is presented to train and evaluate a simple convolutional neural network for radar nowcasting, including self-contained R code.

\section{Introduction}

The history of weather radar, acronym for RAdio Detection And Ranging of objects, dates back to world war II when timely thunderstorm warnings were provided to aviators which opened the possibilities of meteorological science as an essential part of aviation \cite{Doviak1993}. 
Today weather radar resources are useful in a variety of fields including military, nautical, aviation, marine, and biology \cite{Binetti2022}. 
Radar-based remote sensing is a particularly important method in modern meteorology \cite{Fukao2014a}. 
Weather radar is commonly associated with the detection of precipitation and storms, but can also provide data of wind, temperature, turbulence and water vapour \cite{Fukao2014b}. 
Accurate information on rainfall and snowfall directly supports forecasting, flood monitoring, and water resource management.

Nowcasting is defined by the WMO\footnote{World Meteorological Organisation} as "forecasting with local detail, by any method, over a period from the present up to six hours ahead" \cite{WMO2017NowcstingGuidelines}.
Since precipitation changes quickly and can have immediate impacts, nowcast-based warning systems therefore rely on the availability of timely, high-resolution observations. 
Radar plays a crucial role in nowcasting by providing near-real-time measurements of precipitation structure, intensity, and motion, which are essential for detecting hazardous weather and issuing early warnings \cite{Yeary2014}. 
This paper contributes a new UK-based data set to train and evaluate nowcasting methods.

Nowcasting approaches include optical flow methods, physics-based numerical models, spatio-temporal statistical modelling and machine learning methods \cite{Prudden2020}. 
Optical flow methods \cite{Bowler2004, Bechini2017} use a prescribed or inferred velocity field to propagate current radar image data into the future. 
Machine-learning methods are typically based on deep artificial neural networks with many thousands of trainable parameters optimised by gradient descent on a suitable data loss function \cite{shi2015convolutional, Agrawal2019, Ayzel2020, metnet2020, Chen2020, Ravuri2021}. 
Statistical nowcasting methods include spatiotemporal Gaussian processes \cite{battagliola2023new} and vector-autoregressive methods \cite{sigrist2012dynamic, johnson2023bayesian}.
All methods include trainable parameters that are fitted to data to take current radar data as input and output an estimate of radar data in the future.

Nowcasting studies vary in the spatial and temporal characteristics of the radar data used for nowcasting model development. \cite{shi2015convolutional} use Hong Kong radar data from 2011–2013 at 6-minute resolution. \cite{metnet2020} train on continental U.S. data at 1 km and 2 minutes. \cite{Chen2020} use 500 \(\times\) 500-pixel radar patches at 6-minute resolution over four years (2015–2018) from the Czech Republic. \cite{Ayzel2020} use German Weather Service composites at 1 km and 5 minutes. \cite{Ravuri2021} use UK radar composites at 1 km and 5 minutes over two years (2016–2019) covering a 1,536 \(\times\) 1,280 \(km^2\) domain.
The data set developed in this study is of lower spatial and temporal resolution than these to allow for fast training and comparison of alternative nowcasting methods.

Machine learning methods rely on suitable training data, which are usually obtained by sampling data of suitable spatial and temporal extent from a larger archive of available radar data.
Previous authors \cite{shi2015convolutional} observed that under random spatial and temporal sampling an unacceptably large fraction of samples show clear-sky conditions, making them unsuitable to train nowcasting models.
To address this, researchers oversample rainy examples for example by ensuring that at least 80\% of examples have at least one pixel of rain \cite{shi2015convolutional}. 
The statistical properties of such rejection-based radar sampling have not been analysed to the best of our knowledge.




The dataset developed in this study contains a collection of high-resolution radar data examples suitable for training and evaluating optical-flow-based, statistical, and machine-learning-based nowcasting methods. 
The data are provided in a portable plain-text format.
A stratified sampling approach ensures spatially and temporally uniform sampling and a minimal rain-intensity threshold for each sample, to avoid the dominance of clear-sky cases observed under random sampling. 
Additional novelty includes the provision of metadata augmenting each radar sequence with geographical and large-scale meteorological information, as well as R code and download instructions for a reproducible workflow, including functions for working with the Met Office Nimrod binary format. 
Together, the published code and data provide a basis for fast evaluation of nowcasting methods on a standard dataset and for generating related datasets with controlled spatial and temporal sampling characteristics.

\section{Data}

\subsection{The "Nimrod" data archive}

The Met Office Nimrod System generates composite radar reflectivity data covering the UK, Ireland and parts of Western Europe \cite{MetOffice2003}.
Composite radar reflectivity data are assembled from single-site radar scans and undergo quality control before publication.
Spatial resolution of this data is 5km, temporal cadence is 15 minutes.
Radar image data have been regularly updated from 2003 until present, currently totaling 41Gb of data\footnote{as of October 2025}.
Data are regularly updated with a time-lag of about 2 days, making them unsuitable for real-time applications but appropriate for statistical analyses and model evaluation studies.
Alternative radar data sets available from the Nimrod system include spatial resolution of 1km, and spatial extent covering all of Europe.
Radar reflectivity fields are compressed in a custom binary format (the Nimrod file format) and made available for download as daily tar archives.
Nimrod data can be downloaded from the CEDA\footnote{Centre for Environmental Data Analysis, \url{https://www.ceda.ac.uk/}} archive \cite{MetOffice2003} by registered users and used under the Open Government License  \cite{OGLv3}.

Specifically, the data described in this paper are from the CEDA directory 
\begin{tcolorbox}
\begin{verbatim}
/archive/badc/ukmo-nimrod/data/composite/uk-5km
\end{verbatim}
\end{tcolorbox}

A complete archive of binary data files was bulk-downloaded on 2025-10-09, with dates ranging from 2004-04-07 to 2025-10-07.
The download included 7,698 daily files and 724,190 total time stamps at a total download size of 11Gb.

\begin{figure}
    \includegraphics[width=\textwidth]{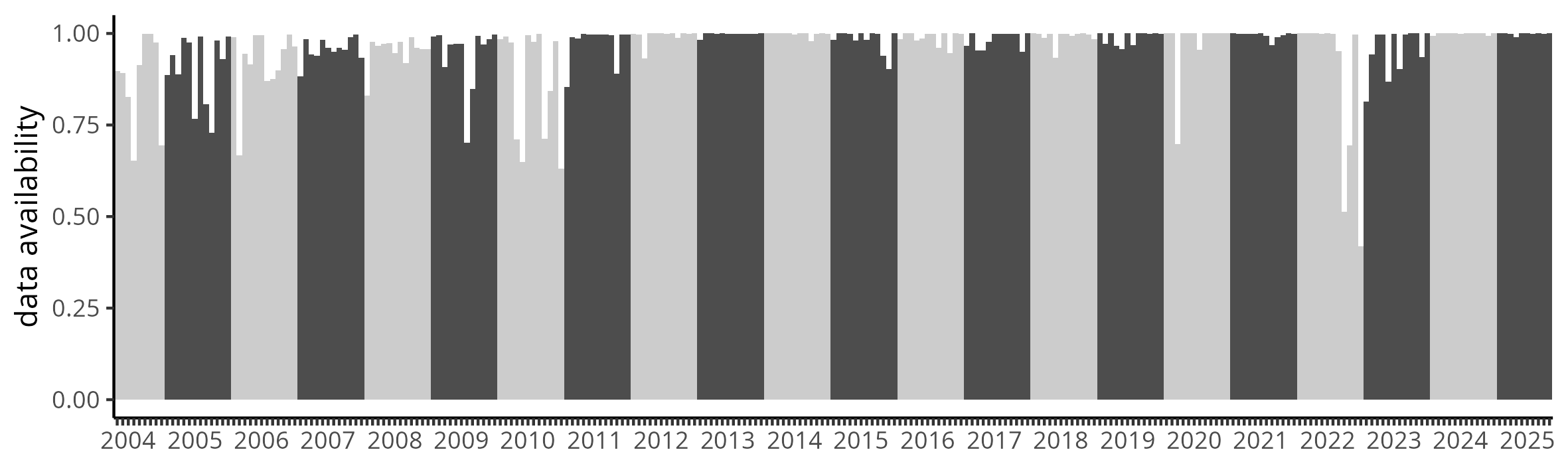}
    \caption{Fraction of available radar fields out of all possible 15-minute time stamps in each month.}
    \label{fig:availability}
\end{figure}

Data are incomplete -- not all 15-minute time slots are available.
Figure \ref{fig:availability} shows fraction of available radar data files per month, relative to all possible 15-minute time slots over the period.
Nimrod radar data are most reliably available from 2011 to 2021 and after 2024.
Exploratory analyses showed no systematic missingness patterns in terms of time of day or day of week.

The Nimrod system includes combination of radar scans at different elevations and from different radar stations for each site.
Quality checking is based on comparison with rain gauge data. 
The Nimrod system has been highlighted repeatedly as a valuable resource, e.g., for diagnosing and analysing severe flood events  \cite{MetOffice2003b}.

\subsection{Processing the Nimrod file format}

The Nimrod format \cite{stone2008nimrod} is a binary file structure in which each file comprises one or more sequential records, each containing a fixed‐length 512‐byte header followed by a data array. 
The header encodes file-specific metadata such as validity times, grid geometry, data type and missing‐value indicators.
This design allows uniform storage of image or model fields (including radar reflectivity) within a common format.
Precipitation intensity data are encoded as integers in units of mm/hr/32, i.e. precipitation intensity of \(q\) mm/hr is encoded by a value \(q \times 32\) rounded to the nearest integer.

Software libraries to extract radar data from the Nimrod file format are available for Fortran, Matlab, IDL and python \cite{nimrodtools}. 
No tools to analyse Nimrod files are widely available for the R statistical programming environment.
The R source code file \verb+nimrod.R+ include R functions written by the authors for data extraction based on the Nimrod file format specification \cite{stone2008nimrod}, as well as several helper functions for bulk-loading, plotting and subsetting data:

\begin{itemize}
\item \verb+nimrod_read_dat(file_name)+ implements the Nimrod file format specifications to read binary data into an integer array, and produce meta data including time stamp, 2D dimensions, and bounding box coordinates in British National Grid coordinates (EPSG:27700, \cite{epsg27700}).
\item \verb+nimrod_read(from, to, by, nimrod_dir)+ is the main function to return a radar image sequence for a given date range (arguments \verb+from+, \verb+to+, \verb+by+) as an R list, wherein each list element refers to one time instance and contains corresponding radar intensity and meta data. The argument \verb+nimrod_dir+ specifies the local top level directory containing the downloaded data.
\item \verb+nimrod_apply_bbox(img, xlim, ylim)+ cuts out a rectangular region from \verb+img+ based on latitudinal and longitudinal limits given as 2-vectors \verb+xlim+ and \verb+ylim+
\item \verb+nimrod_apply_bbox2(img, center, nx, ny)+ cuts out a rectangular region from \verb+img+ based on 2-vector \verb+center+ and numbers of columns and rows of the target output
\item \verb+nimrod_plot(...)+  and \verb+nimrod_animate(...)+ visualise images and sequences.
\item  \verb+nimrod_coast.csv+ is a comma-separated values (csv) file containing UK coast line data used in plots and animations, obtained from Natural Earth data \cite{naturalearth} via the R package \verb+rnaturalearth+ \cite{rnaturalearth}.
\end{itemize}

The following R code examples illustrate the intended usage.

\begin{tcolorbox}
\begin{verbatim}
# R code #
source('nimrod.R')
loc = '~/archive/badc/ukmo-nimrod/data/composite/uk-5km'
\end{verbatim}
\end{tcolorbox}

The directory stored in \verb+loc+ must exist in the local file system.
The following code assumes that \verb+loc+ contains the year-specific subdirectory \verb+2025/+ and the daily radar file for the requested date is available in that directory.

\begin{figure}
    \includegraphics[width=\textwidth]{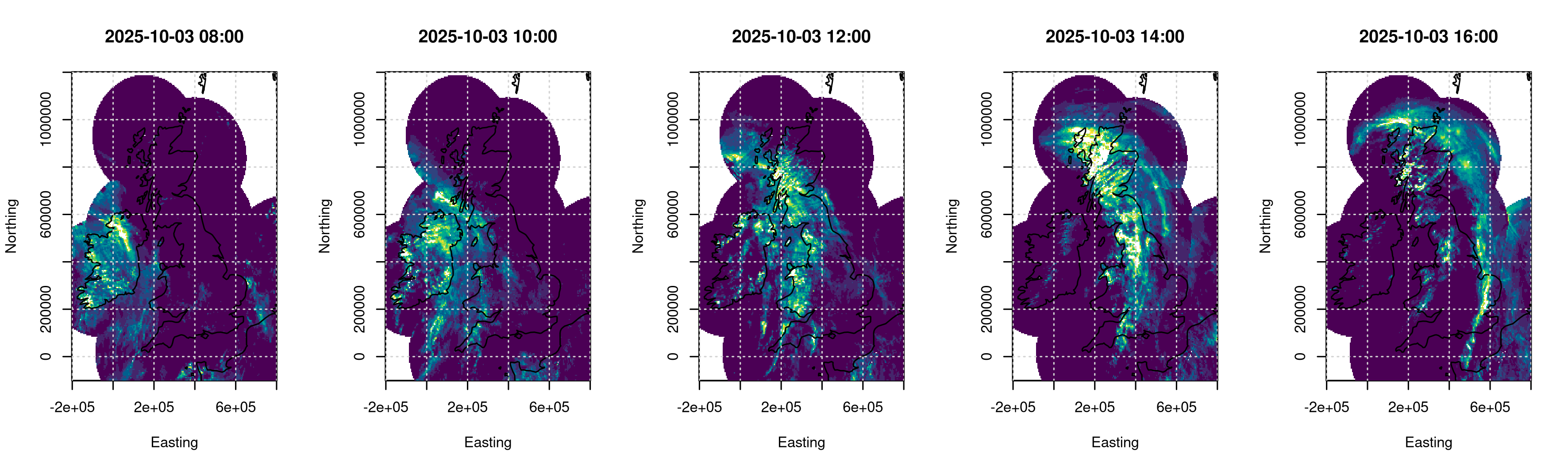}
    \caption{Example plot of Storm "Amy" moving over the UK on 2025-10-03.}
    \label{fig:storm_amy}
\end{figure}

\begin{tcolorbox}
\begin{verbatim}
# R code #
imgs =  nimrod_read(from = '2025-10-03 08:00', 
                    to   = '2025-10-03 16:00', 
                    by   = '2 hour', 
                    nimrod_dir = loc)
                    
imgs = nimrod_apply_bbox(imgs, xlim=c(-2e5, 8e5), 
                               ylim=c(-1e5,12e5))
\end{verbatim}
\end{tcolorbox}
The following command plots the sequence of images shown in Fig.~\ref{fig:storm_amy}:
\begin{tcolorbox}
\begin{verbatim}
# R code #
coast = read.csv('nimrod_coast.csv')
layout(t(1:5))
for (i in 1:5) {
  nimrod_plot(imgs[[i]], zlim=c(0,200))
  lines(coast)
  grid()
}
\end{verbatim}
\end{tcolorbox}
On X11-based systems, an animation of the same sequence is produced with the following code:
\begin{tcolorbox}
\begin{verbatim}
# R code #
nimrod_animate(imgs, coast=coast, zlim = c(0,200))
\end{verbatim}
\end{tcolorbox}

\section{Sampling}

\subsection{Sampling protocol}

This section describes how a data set of complete, spatially and temporally homogeneous and intensity-thresholded radar sequences was sampled from the larger Nimrod data set.
The R code file \verb+sample.R+ contains the complete code.
To run the sampling code, the following parameters can be set at the beginning of the script, with values used to generate the published data set shown in parentheses:
\begin{itemize}
    \item random number generator seed (1)
    \item number of samples in final data set (1000)
    \item sequence length per sample (20)
    \item spatial extent per sample (40 \(\times\) 40)
    \item spatial sampling domain $[0,6 \cdot 10^5] \times [0, 8 \cdot 10^5]$ (EPSG:27700)
    \item sampling time period (2011-01-01 00:00 -- 2020-12-31 23:59)
    \item total precipitation rejection threshold (\(10^5\))
\end{itemize} 
Even though the UK is notorious for rainy weather, unconditional random sampling results in a large number of clear sky conditions. 
We concur with previous authors \cite{shi2015convolutional} that such cases should not be contained in a data set intended to train precipitation nowcasting models.
The rationale to exclude clear sky or low precipitation situations to generate the sample is that under non-rainy conditions a nowcasting model would usually not be applied.
To avoid inclusion of low precipitation cases, we reject radar sequences with total aggregate precipitation intensity below the specified threshold.

We found that uniform spatial and temporal sampling, followed by rejection of sequences with total precipitation below the threshold, resulted in a systematic bias of sample locations towards the (wetter) Northwestern part of the country. 
To maintain spatial homogeneity under rejection sampling, we
use spatially stratified uniform sampling, wherein the full spatial domain is partitioned into 12 equal-sized areas (strata) of size 200km x 200km and spatially and temporally uniform random sampling followed by threshold-based rejection was performed separately within each stratum. 
The in-stratum sample sizes were set to obtain a sample of the desired total sample size that is balanced across strate.

After setting the sampling parameters the data set is generated by iterating over the following sampling protocol, separately within each stratum:

\begin{enumerate}
\item Sample a (uniformly distributed) random date and hour of day within the time sampling period to set the start time of the sampled sequence.
\item Sample a (uniformly distributed) random location from the current spatial sampling stratum to set the center point of the sampled sequence.
\item Extract radar reflectivity data of the required spatial extent and length for the sampled time point and center location
\item If the data contains any missing values or if the total aggregated precipitation amount is below the threshold, reject the sample and go to~1.
\item Store the sampled sequence.
\item Go to 1. or stop if the target number of samples has been reached.
\end{enumerate}

\begin{figure}[!htbp]
\begin{minipage}{.45\textwidth}
\includegraphics[width=\textwidth]{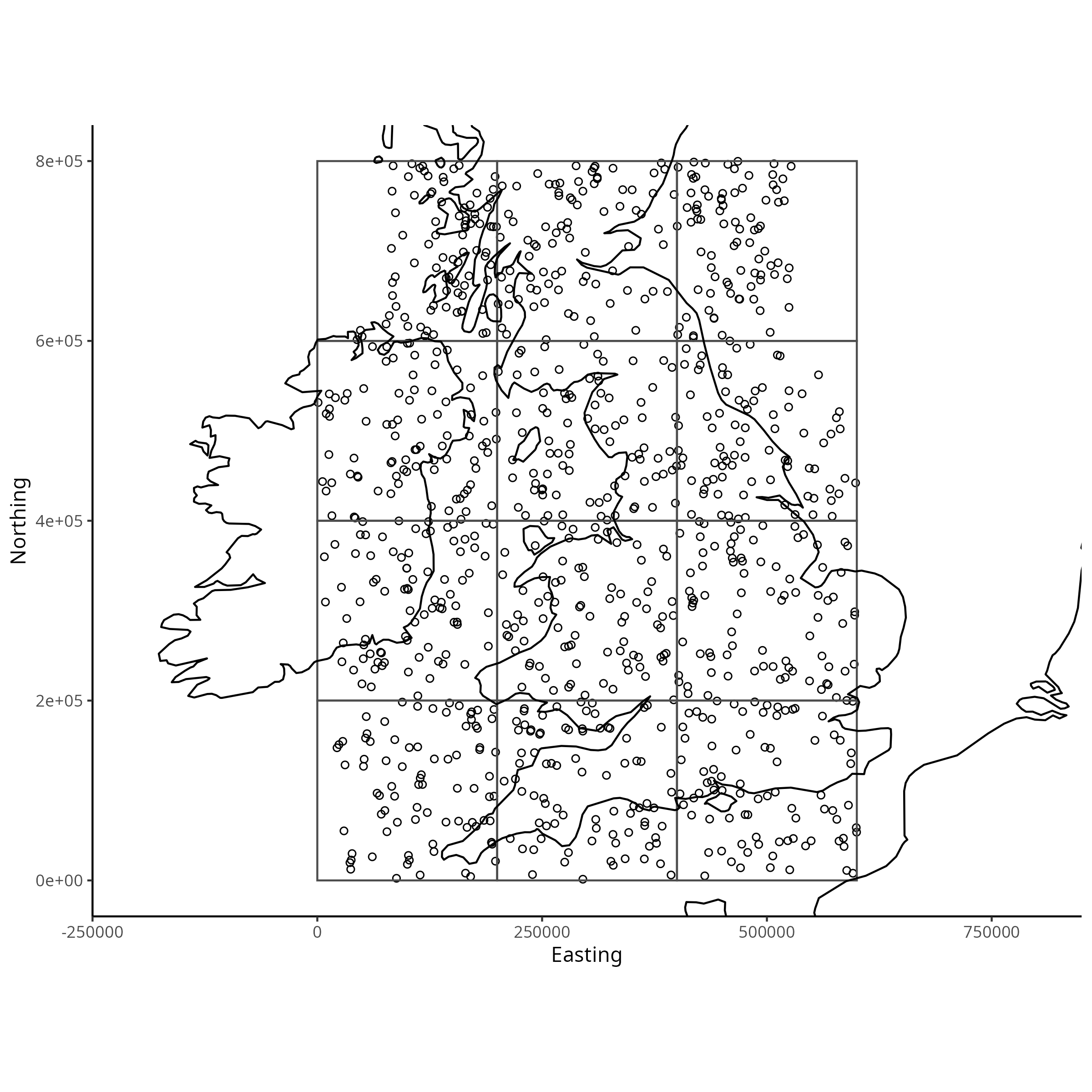}
\end{minipage}%
\begin{minipage}{.55\textwidth}
\includegraphics[width=\textwidth]{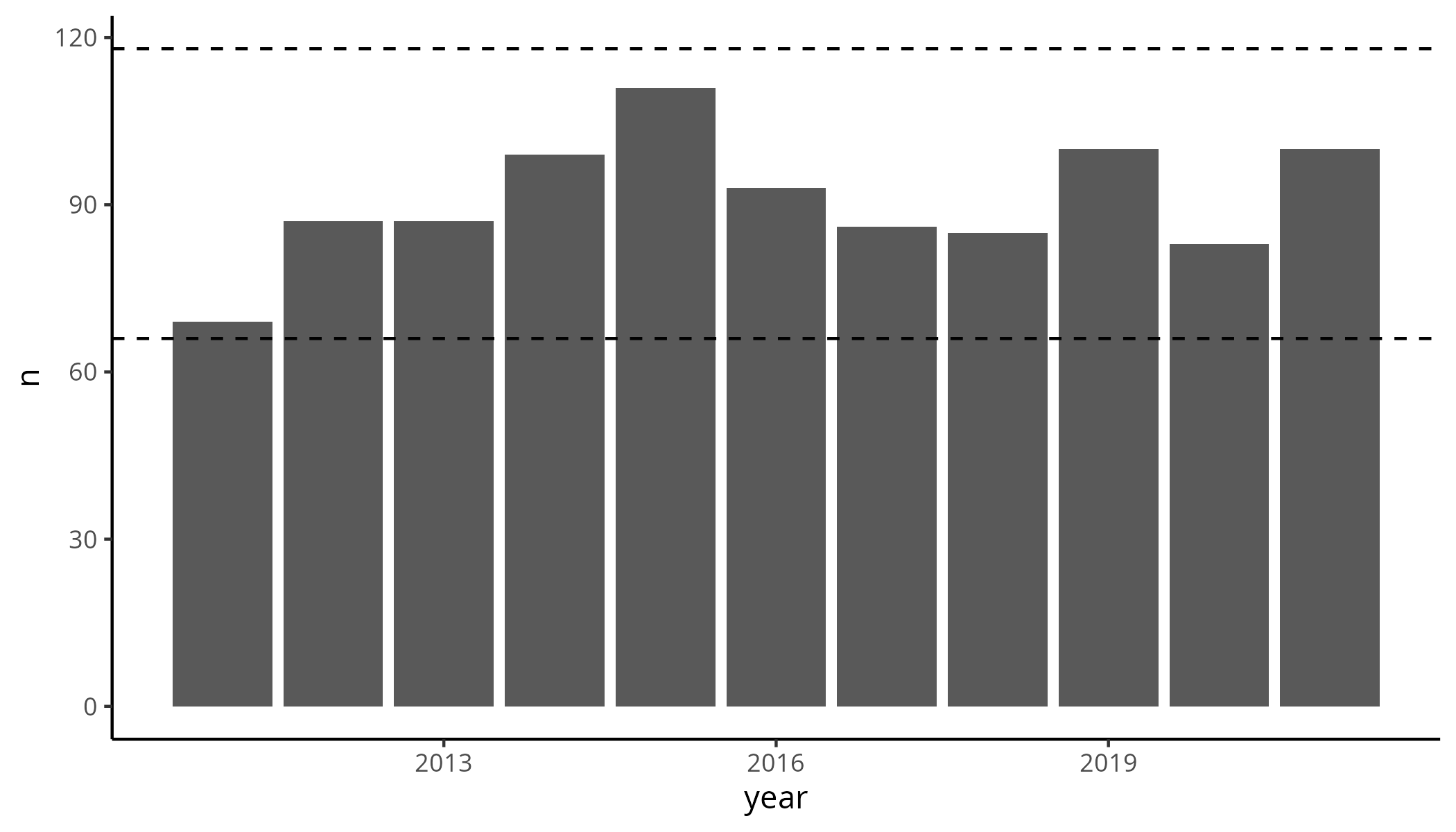}
\end{minipage}%
\caption{Left: Sampling locations and spatial strata. Right: Number of samples per year compared to 95\% simultaneous probability interval of the Binom(1000, 1/11) distribution (Bonferroni corrected).}
\label{fig:sampls-space-time}
\end{figure}

Figure~\ref{fig:sampls-space-time} shows the homogeneous distributions of spatial sampling locations and years.
We note minor artifacts in the Northern corners of the sampling domain due to limited availability of radar data of the required spatial extent.
We consider these acceptable.
The year-on-year variability of sample size is consistent with uniform random sampling across years as shown by the Bonferroni-corrected 95\% uncertainty intervals.

\begin{figure}[!htbp]
    \includegraphics[width=\textwidth]{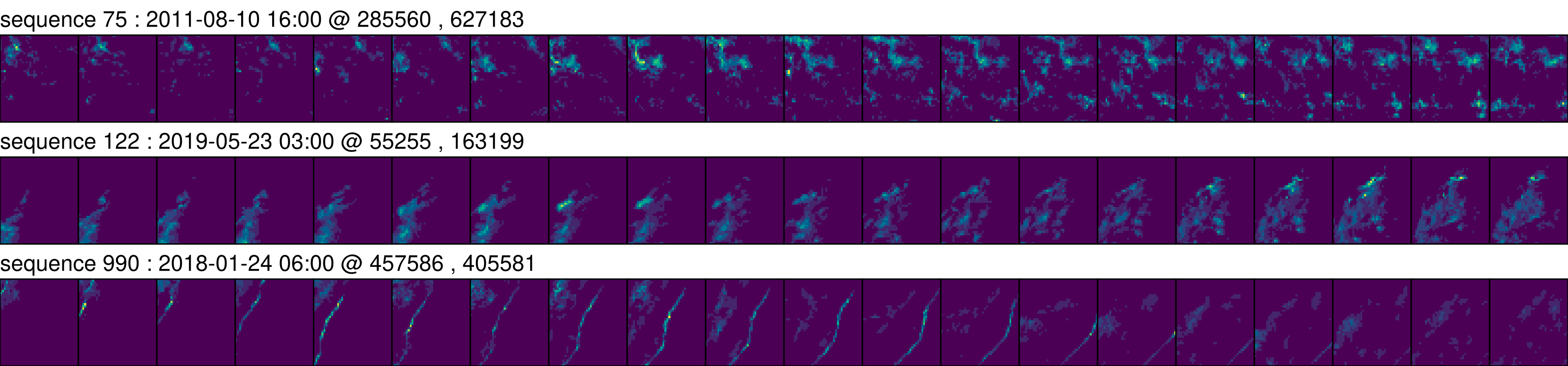}
    \caption{Sample sequences from different locations and seasons. Brighter colours indicate higher precipitation rate. Date and time of first frame and coordinates of center point are given in captions.}
    \label{fig:sample-seq}
\end{figure}

\begin{figure}[!htbp]
    \centering
    \includegraphics[width=\linewidth]{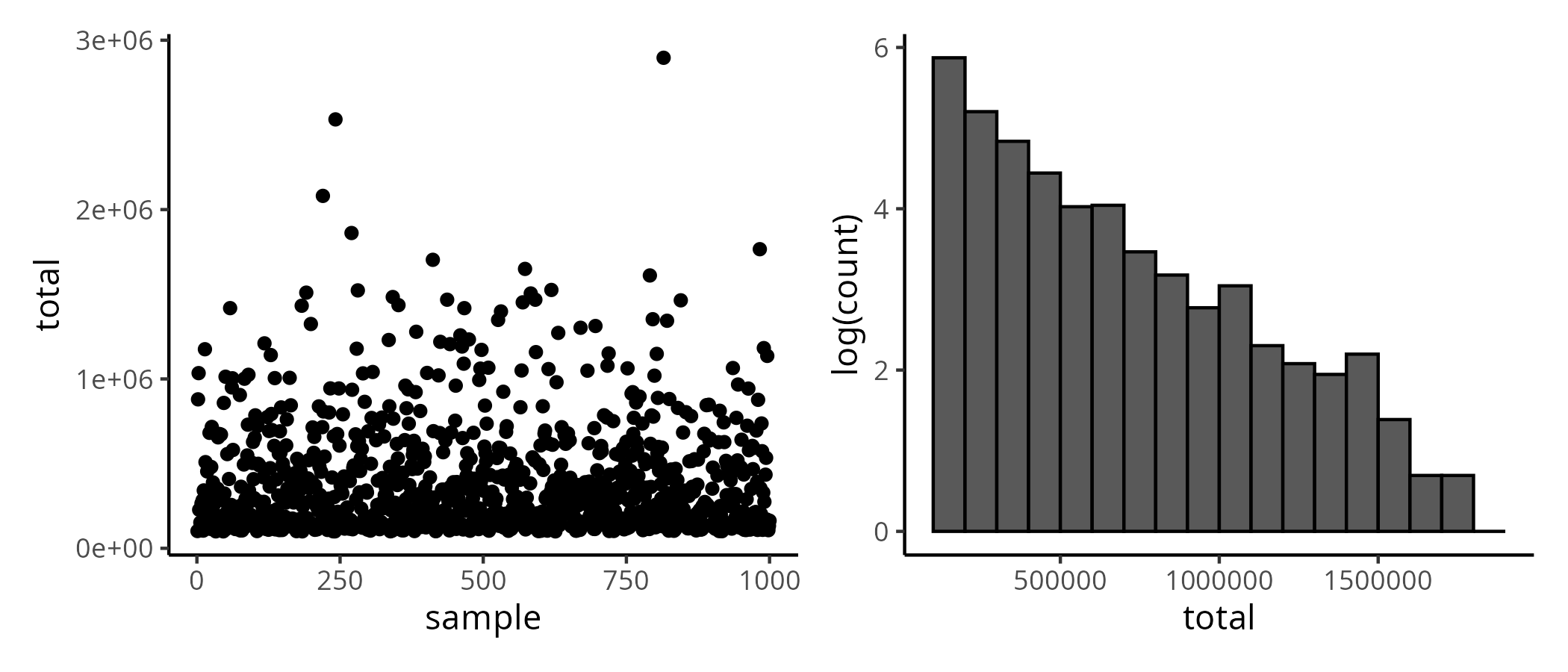}
    \caption{Total precipitation aggregates per sampled sequence, and histogram in log-linear scales.}
    \label{fig:totals-hist}
\end{figure}

Figure~\ref{fig:sample-seq} shows three examplary radar sequences from the full data set.
Figure~\ref{fig:totals-hist} summarises total aggregated preciptitation amounts per sampled sequence.
The histogram in log-linear scales shows that aggregates approximately follow an exponentially distribution.
The memorylessness of the exponential distribution ensures that further thresholding with higher threshold values would reduce the sample size but would not alter the shape of the distribution.

\subsection{Additional features}\label{sec:additional}

Each radar sequence in the data set was further augmented with a set of geographically and meteorologically relevant scalar variables pertaining to the start of the sequence. 
These auxiliary features provide additional context that can be used for stratified analyses or for conditioning nowcasting models on additional information. 
The additional variables include terrain height statistics, summaries of horizontal and vertical winds, and categorical storm-type labels as detailed in this section.

Elevation data for each spatial sampling window was obtained from the GEBCO data set \cite{GEBCO_2025_Grid} and the following elevation summaries were calculated for each sample:
\begin{itemize}
    \item mean elevation within the sample region
    \item maximum elevation within the sample region
    \item fraction of points in the sample region with zero elevation 
\end{itemize}
Mean and maximum elevation characterise elevation profile, and zero fraction can be used as a proxy for water coverage.

Wind data at 850hPa pressure level were obtained from hourly ERA5 reanalysis data \cite{hersbach2020era5}, for start time and spatial extent of each sequence. 
We calculated scalar summaries of the local wind fields, specifically
\begin{itemize}
    \item U850: zonal wind velocity at 850hPa in m/s averaged over the sample region
    \item V850: meridional wind velocity at 850hPa in m/s averaged over the sample region
    \item D850: horizontal divergence at 850hPa in $10^{-5}$/s averaged over the sample region
\end{itemize}
Zonal and meridional wind speed can be used as proxies of large scale horizontal motion.
Horizontal divergence is provided as a diagnostic of lower-tropospheric convergence, a key factor in convective development.

\begin{figure}[!htbp]
    \centering
    \includegraphics[width=0.5\linewidth]{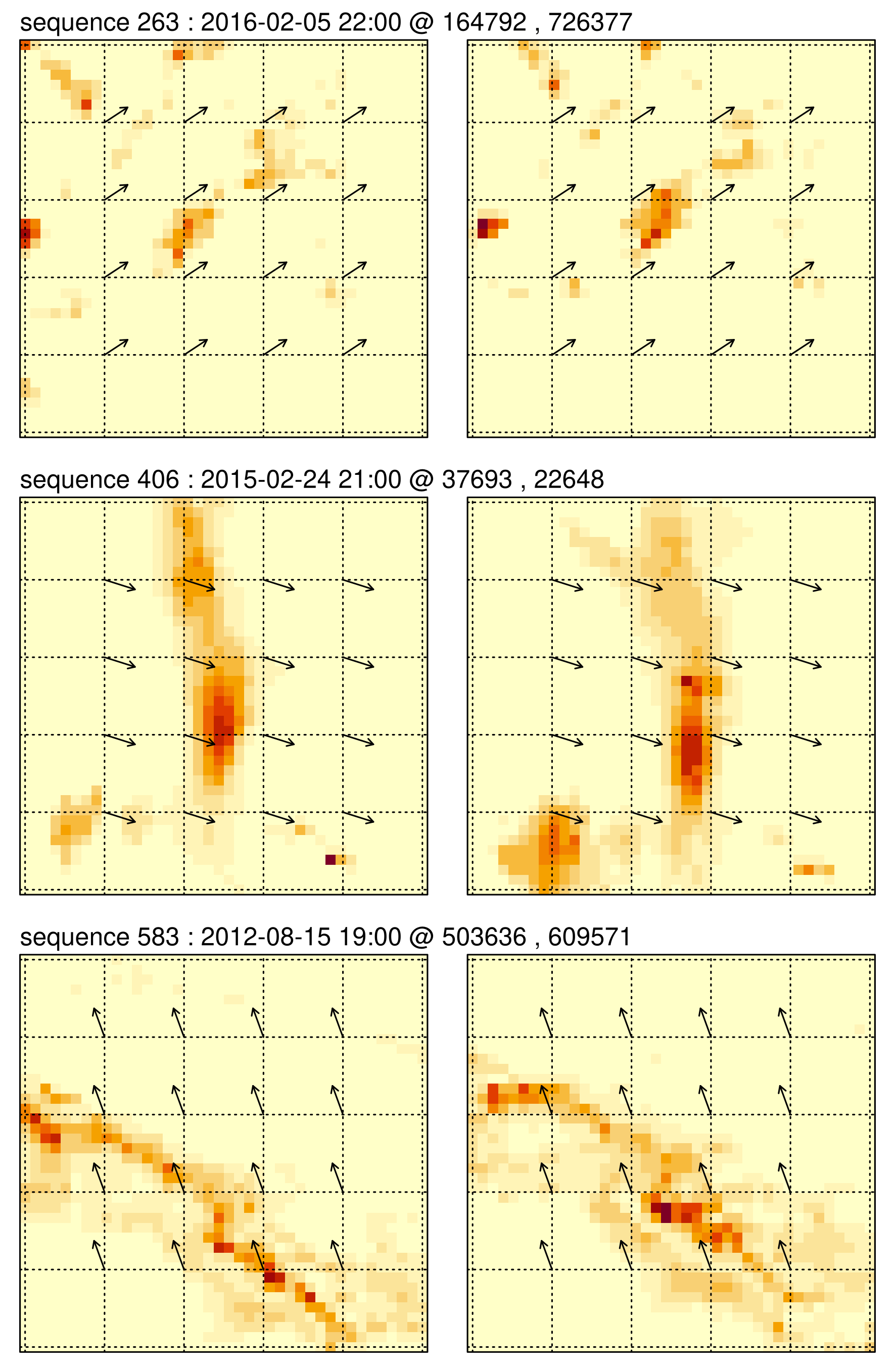}%
    \includegraphics[width=0.5\linewidth]{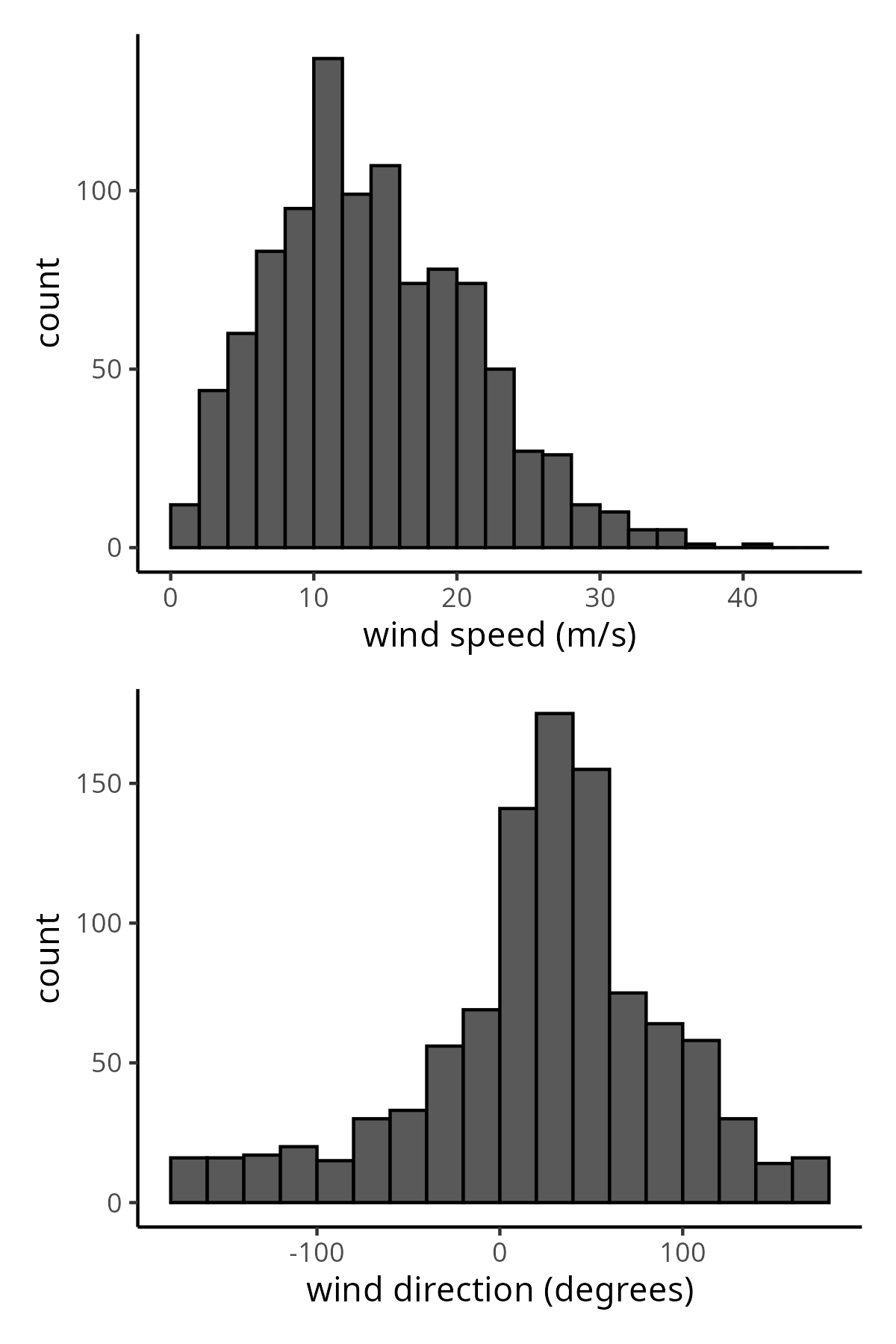}
    \caption{Left: Comparison of consecutive images in Nimrod radar sequences with corresponding ERA5 region-averaged wind vector (scaled to units of 5 km / 15 minutes). Right: Summary histograms of wind speed and direction over 1000 samples.}
    \label{fig:wind-validation}
\end{figure}
\FloatBarrier

Storm-type information was obtained from the ERA5-based dataset by \cite{CattoDowdy2023},  which classifies synoptic-scale weather systems over the North Atlantic and Europe into eight categories covering all possible combinations of cyclone, fronts, and thunderstorms by assigning them to integers as follows:
\begin{itemize}
    \item 0 - no storm type
    \item 1 - cyclone only
    \item 2 - front only
    \item 3 - thunderstorm only
    \item 4 - cyclone + front
    \item 5 - cyclone + thunderstorm
    \item 6 - front + thunderstorm
    \item 7 - cyclone + front + thunderstorm
\end{itemize}
Stormtypes are temporal aligned with sample time by using the record that is valid at or immediately before the sequence start time (ensuring no use of information from the future).
Spatial matching is ensured by choosing the nearest grid point to the sequence's centre.
Storm type data are currently only available until 2018 – temporal coverage can be extended in future data releases.
When no exact or nearby storm record exists for the given sample time or location the storm type label is set to \texttt{NA}.

\subsection{Sample files structure}

The generated sample of radar sequences is stored in plain text files for portability across platforms and programming environments.
A sample file containing 1000 sequences of length 20 with spatial extent 40x40 is structured as follows:
\begin{tcolorbox}
\begin{verbatim}
# 2014-11-05 02:00 198141.413461417 21397.4663056433
0 0 0 0 0 0 0 0 2 7 0 0 0 5 0 ... <1600 integers per line>
... <19 more lines>
# 2012-10-05 18:00 406519.159395248 170669.694337994
... <20 lines of 1600 integers per block>
...
... <total of 1000 blocks of 21 lines>
...
\end{verbatim}
\end{tcolorbox}
Each radar sequence is stored as a block of \(C+1\) rows where \(C\) is the sequence length (here \(C=20\)).
The first line in each block contains time stamp of the first image in the sequence and spatial coordinates (in EPSG:27700) on which the sequence is centered.
The remaining \(C\) lines of data in each block represent vectorised image matrices (row-major format, upper-right origin), as a (whitespace-separated) sequence of \(W \times H\) integers (here \(W\times H = 40 \times 40\)).
A file with sample size \(B\) contains \(B\) such blocks, or a total of \(B \times (C+1)\) lines of text.

The following exemplary R code shows how to load all data into a 4-dimensional array of dimension \((B,H,W,C)\), by removing comment lines, serialising into a vector of integers, casting that sequence into an array of dimension \((W,H,C,B)\) and permuting dimensions to obtain the required final shape:

\begin{tcolorbox}
\begin{verbatim}
# R code #
B = 1000; W = 40; H = 40; C = 20
arr = 
  readLines('NimrodMLdata-1000x40x40x20-seed1.dat') |>
  grep(pattern='#', invert=TRUE, value=TRUE) |>
  strsplit(' ') |>
  unlist() |>
  as.integer() |>
  array(dim=c(W, H, C, B)) |>
  aperm(c(4,2,1,3))
\end{verbatim}
\end{tcolorbox}

Python/Numpy code to do the same is as follows:

\begin{tcolorbox}
\begin{verbatim}
# python code #
import numpy as np
W, H, C, B = 40, 40, 20, 1000
data = np.loadtxt('NimrodMLdata-1000x40x40x20-seed1.dat',
                  comments='#', dtype=int)
arr = data.ravel().reshape(B, C, H, W).transpose(0, 2, 3, 1)
\end{verbatim}
\end{tcolorbox}

Note that R and Python use different linearisation orders (column-major vs. row-major). 
Directly casting to an array of the final dimensions would scramble the data, so we have to load into the native order first and then permute axes.

Meta information and additional (scalar) features for each sequence are stored in the file
\verb+NimrodMLdata-1000x40x40x20-seed1.csv+
including date and time of first frame, coordinates of region center in EPSG:4326 and EPSG:27700 as well as the additional features described in section \ref{sec:additional}.

The column names and first few lines of the meta data file are shown in Table \ref{tab:meta}.
\begin{table}
{\tiny
\begin{tabular}{llrrrrrrrr}
\toprule
date & time & easting & northing & longitude & latitude & elev & elev\_max & elev\_frac0 & ... \\
\midrule
2014-11-05 & 02:00 & 198141.4 & 21397.47 & -4.82 & 50.06 & 36.98 & 598 & 0.73 & ...\\
2012-10-05 & 18:00 & 406519.2 & 170669.69 & -1.91 & 51.43 & 95.15 & 791 & 0.12 & ...\\
2018-11-07 & 06:00 & 360169.3 & 132246.46 & -2.57 & 51.09 & 91.29 & 796 & 0.29 & ...\\
2018-11-29 & 06:00 & 520552.4 & 189038.86 & -0.26 & 51.59 & 67.56 & 287 & 0.10 & ...\\
2016-11-22 & 09:00 & 241396.7 & 434119.31 & -4.41 & 53.78 & 41.99 & 1005 & 0.75 & ...\\
\bottomrule
\end{tabular}\\[1em]

\begin{tabular}{rrrrr}
\toprule
... & u850 & v850 & d850 & stormtype\\
\midrule
... & 5.74 & -11.01 & 1.06 & 3 \\
... & 9.36 & 8.54 & -1.30 & 2 \\
... & 13.17 & 13.08 & -0.94 & 7 \\
... & -9.72 & -5.21 & 1.28 & 4 \\
...  & 8.52 & 18.84 & 1.02 & 4 \\
\bottomrule
\end{tabular}
}
\caption{First 5 rows of meta data accompanying radar samples.}
\label{tab:meta}
\end{table}

\section{Case study: Training a Deep Convolutional Neural Network}\label{sec:cnn}

The purpose of the sampled sequences is to fit and test radar nowcasting models.
As a case study we show self-contained code to fit and analyse a simple convolutional neural network (CNN) for next-frame prediction.
Initially, we randomly sample 75\% of cases as training data, and select the first 4 images in each sequence as inputs, and the 5th image as the prediction target. 
\begin{tcolorbox}
\begin{verbatim}
# R code #
i_train = sample(B, floor(B * 0.75))
x_train = arr[i_train, , , 1:4, drop=FALSE]
y_train = arr[i_train, , , 5,   drop=FALSE]
# dim(x_train) is (750, 40, 40, 4)
# dim(y_train) is (750, 40, 40, 1)
\end{verbatim}
\end{tcolorbox}
We use the \verb+keras3+ R package \cite{keras3} to initialise a deep convolutional neural network with 2 hidden 2d convolution layers with kernel size 3 and ReLU activation function.
\begin{tcolorbox}
\begin{verbatim}
# R code #
library(keras3)
model = keras_model_sequential(input_shape = c(40,40,4))
model |> 
  layer_conv_2d(filters = 32, kernel_size = c(3,3), 
                padding = 'same', activation = 'relu') |>
  layer_conv_2d(filters = 16, kernel_size = c(3,3), 
                padding = 'same', activation = 'relu') |>
  layer_conv_2d(filters = 1, kernel_size = c(3,3), 
                padding = 'same', activation = 'relu') 
\end{verbatim}
\end{tcolorbox}
This deep CNN has 5,953 trainable parameters.
Using \verb+padding="same"+ in each convolution layer and \verb+filters=1+ in the last layer ensures that the model output is of shape (40,40,1).
The number of hidden layers, the number of filters per hidden layer and kernel size are hyperparameters.
The model is compiled to use mean-squared error loss and the Adam optimiser, and train for 50 epochs with batch size 64:
\begin{tcolorbox}
\begin{verbatim}
# R code #
model |> compile(loss = 'mse', optimizer = 'adam')
model |> fit(x_train, y_train, batch_size = 64, epochs = 50)
\end{verbatim}
\end{tcolorbox}
For out-of-sample evaluation, we sample the test data and calculate MSE of the trained model
\begin{tcolorbox}
\begin{verbatim}
# R code #
i_test = setdiff(1:B, i_train)
x_test = arr[i_test, , , 1:4, drop=FALSE]
y_test = arr[i_test, , , 5,   drop=FALSE]
pred_test = model(x_test) |> as.array()
mean((pred_test - y_test)^2) # 1006.575 
\end{verbatim}
\end{tcolorbox}
For benchmarking, we also calculate the MSE of the no-change persistence prediction:
\begin{tcolorbox}
\begin{verbatim}
# R code #
pred_test_ref = x[, , , 4, drop=FALSE]
mean((pred_test_ref - y_test)^2) # 1768.452
\end{verbatim}
\end{tcolorbox}
In our example run, the CNN reduced the MSE from 1768.5 to 1006.6, indicating a significant improvement over the no-change reference prediction. 
Similar improvements of the CNN were seen when comparing against the all-zero forecast, and against the climatological mean forecast.
The code to fit the same CNN using python, Numpy and keras, producing similar performance metrics, is in Appendix~\ref{app:cnn-python}.
\begin{figure}
    \centering
    \includegraphics[width=0.7\linewidth]{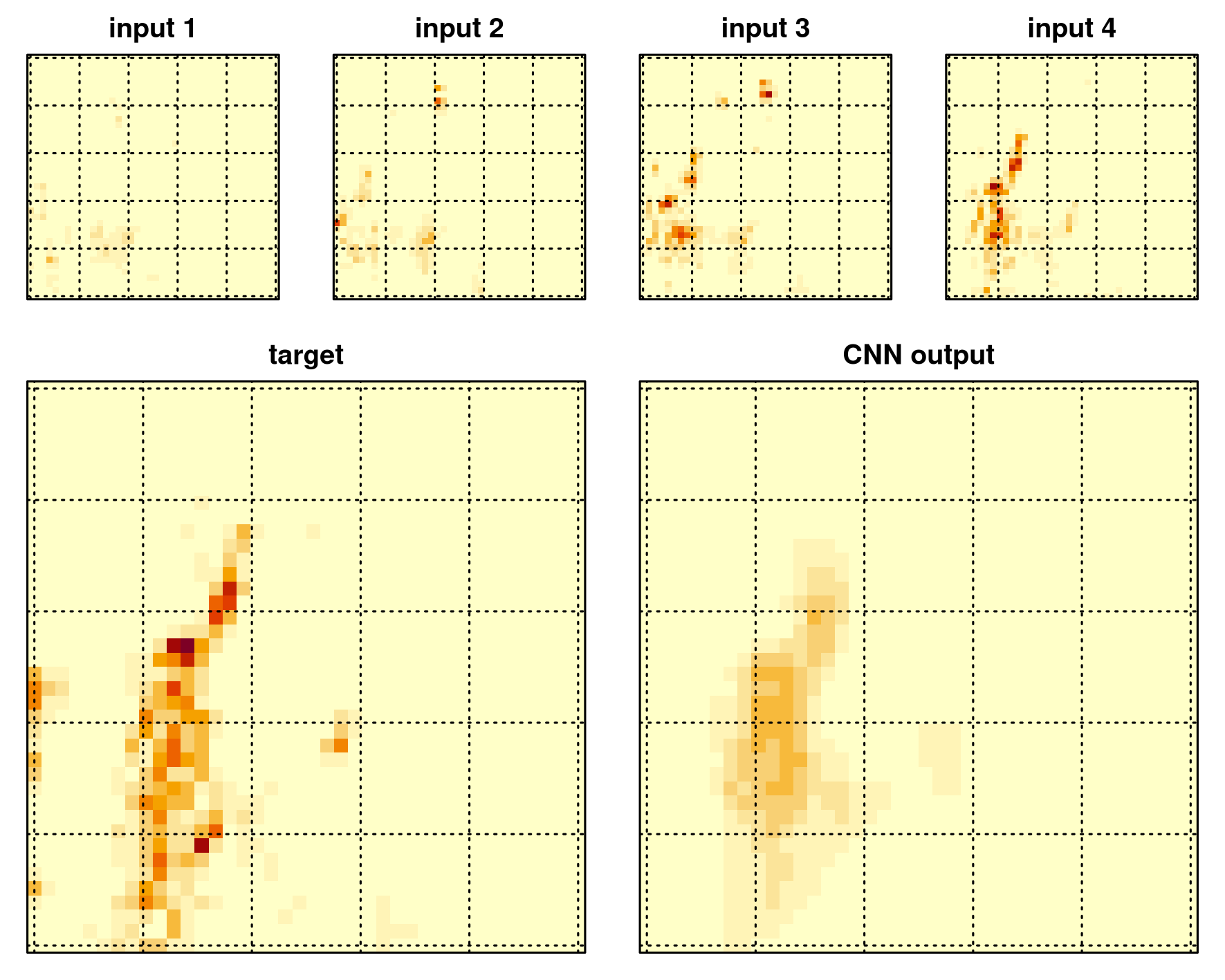}
    \caption{Comparison of CNN input, CNN output, and target.}
    \label{fig:cnn-output}
\end{figure}

Figure \ref{fig:cnn-output} shows a specific example of model output. 
It illustrates that the CNN prediction is too smooth compared to the target, which has been noted by other authors before \cite{li2021MSDM,han2023key}.
However, we also note that the main mass is propagated in the correct direction, showing evidence that the model has learned to exploit advective transport for prediction.
The CNN output could be improved by a post-processing step that recovers small scale features, e.g. by reverse-diffusion modelling \cite{Mardani2025CorrDiff}, or choosing a U-Net architecture which recovers small spatial scales by skip connections \cite{Ayzel2020} but testing such improvements is beyond the scope of this paper.

\section{Discussion and Conclusion}

Radar reflectivity data has well-documented artifacts, non-homogeneities and other shortcomings, due to gaps and overlaps in coverage, variations in data format and processing standards, and data usage restrictions \cite{Agrawal2019, Saltikoff2019}
The present paper made no attempt to correct or filter any artifacts. 
Only "raw data" is provided here, subject only to any quality control applied by the data provider.

The provided data is designed to allow training models of different complexity, and their comparison, without access to HPC facilities.
The simple CNN used in section \ref{sec:cnn} trains in only a few seconds on a standard consumer laptop (Intel i7 CPU, 32Gb RAM, no GPU).
The data size is therefore suitable for exploratory studies of model architectures and hyperparameters.

Nowcasting applications can vary in spatial extent, model choice, regional focus, an these factors should be reflected in the training data set.
By providing source code for data extraction and sampling, we enable adaptions of the data set to different application cases.
Note, however, that generating new data sets requires access to the Nimrod data archive.

The additional large-scale meteorological factors can be used for stratified performance analysis of a trained nowcasting model.
For example, the individual case shown in Fig.~\ref{fig:cnn-output} might look successful not because the model has learned to adapt to the current direction of flow, but because the model has learned to advect all images into the dominant direction of flow shown in Fig.~\ref{fig:wind-validation}.
A performance analysis stratified by current wind direction could test such a hypothesis.

The additional features can also be used as additional real-time features provided as inputs to the nowcasting model.
We expect the information contained in wind and stormtypes to be exploitable to improve model performance.
In practice, this could be achieved for example by using divergence or storm type information to modulate stochastic innovations in autoregressive nowcasting models, or providing current horizontal wind direction and speed as input to a physics-informed machine learning model.

\section{Acknowledgements}

VA was supported by the UK Engineering and Physical Sciences Research Council (ESPRC) Doctoral Training Partnership grant EP/W523859/1.

\bibliographystyle{plain}
\bibliography{main}

\appendix

\section{CNN python code}\label{app:cnn-python}

\begin{tcolorbox}[breakable]
\begin{verbatim}
# python code #
import numpy as np
from tensorflow.keras.models import Sequential
from tensorflow.keras.layers import Conv2D, Input

# convert arr to float32 assumed by keras
arr = arr.astype(np.float32, copy=False)

# training data
i_train = np.random.choice(B, size=B*3//4, replace=False)
x_train = arr[i_train, :, :, 0:4]
y_train = arr[i_train, :, :, 4:5]

# initialise model
model = Sequential([
    Input(shape=(40, 40, 4)),
    Conv2D(filters=32, kernel_size=(3, 3), 
           padding='same', activation='relu'),
    Conv2D(filters=16, kernel_size=(3, 3), 
           padding='same', activation='relu'),
    Conv2D(filters=1,  kernel_size=(3, 3), 
           padding='same', activation='relu')
])

# compile and fit model
model.compile(loss='mse', optimizer='adam')
model.fit(x_train, y_train, batch_size=64, epochs=50)

# calculate test MSE
i_test = np.setdiff1d(np.arange(B), i_train)
x_test = arr[i_test, :, :, 0:4]
y_test = arr[i_test, :, :, 4:5]
pred_test = model.predict(x_test)
np.mean((pred_test - y_test) ** 2)

# MSE of the persistence prediction
pred_test_ref = x_test[:, :, :, 3:4]
np.mean((pred_test_ref - y_test) ** 2)
\end{verbatim}
\end{tcolorbox}

\end{document}